\documentclass[floatfix,%
 reprint,
superscriptaddress,
 amsmath,amssymb,
 aps,
pra, onecolumn
]{revtex4-2}
\usepackage{graphicx}
\usepackage{dcolumn}
\usepackage{bm}
\usepackage{hyperref}
\usepackage{amsfonts}
\usepackage{xcolor}
\usepackage{comment}
\usepackage{booktabs}
\usepackage[normalem]{ulem}
\usepackage{subfigure}
\usepackage{makecell}
\usepackage{braket}
\usepackage{float}

\usepackage{ulem} 

\def\comment#1{}
\def\bra#1{\mathinner{\langle{#1}|}}
\def\ket#1{\mathinner{|{#1}\rangle}}

\def\beq{\begin{equation}}
\def\eeq{\end{equation}}
\def\bea{\begin{eqnarray}}
\def\eea{\end{eqnarray}}

\def\l{\left}
\def\r{\right}

\def\p{\textbf{p}}
\def\q{\textbf{q}}

\def\comment#1{}

\begin{document}

\title{Dephasing through Bremsstrahlung emission: insights from quantum Boltzmann equation}

   \author{Hassan Manshouri}
   \email[]{h.manshouri@ph.iut.ac.ir}
   
   \affiliation{Department of Physics, Isfahan University of
   	Technology, Isfahan 84156-83111, Iran}
   \affiliation{Quantum Technology Research Group, Isfahan University of Technology, Isfahan 84156-83111, Iran}
   
   \author{Moslem Zarei}
   \email[]{m.zarei@iut.ac.ir}
   
   \affiliation{Department of Physics, Isfahan University of Technology, Isfahan
   	84156-83111, Iran}
   \affiliation{Quantum Technology Research Group, Isfahan University of Technology, Isfahan 84156-83111, Iran}

   \date{\today}


\begin{abstract}

We investigate decoherence mechanisms in open quantum systems using quantum field theory techniques and the quantum Boltzmann equation. Specifically, we focus on decoherence through Bremsstrahlung emission, a fundamental process in quantum electrodynamics leading to coherence loss. By applying quantum field theory techniques and quantum Boltzmann equation, we model the fermion-photon interaction in the Stern-Gerlach interferometer and analyze the induced dephasing factor.
Our approach offers significant advancements in understanding decoherence and its potential applications in quantum sensing and atomic interferometry. We demonstrate the accuracy of our method by comparing results to classical Bremsstrahlung.

\end{abstract}

\maketitle

\section{Introduction}
High precision advanced quantum
atom interferometry devices have become indispensable tools for monitoring fundamental physical phenomena \cite{dimopoulos2007,lan2013,muller2007,stray2022}. Unlike traditional interferometers that use light, atom interferometry utilizes atoms to measure various physical quantities, including time, gravity, and acceleration \cite{Margalit_2021}. In the atom interferometry setup, a cloud of atoms is split into two or more paths using laser pulses. Subsequently, the paths recombine, and the resulting interference pattern of the atoms reveals the measured physical quantity. Atom interferometry boasts several advantages over conventional measurement techniques, such as high precision and accuracy, low noise, and the ability to measure non-classical states of matter. These versatile atom interferometry devices find applications in diverse fields, ranging from precision metrology to gravity gradiometry and inertial navigation \cite{hu2017}. Moreover, they have played a crucial role in testing fundamental physics theories, including experiments related to dephasing and decoherence due to electromagnetic interactions \cite{Schut_2024,Yoshimura_2024}, the equivalence principle of general relativity \cite{Asenbaum_2020} and investigations into the nature of dark matter \cite{Du_2022,Hajebrahimi_2023}. 

In recent years, atom interferometry, an idea conceived after the discovery of the Stern-Gerlach (SG) effect in 1922 \cite{Gerlach1922}, has garnered significant attention. The SG interferometer utilizes a magnetic gradient to separate particles with different spin projections into distinct momentum states, followed by their recombination to create a superposition state of spin and momentum \cite{boustimi2000,dimopoulos2007,clade2009}. Initially, SG interferometry was deemed impractical due to the Humpty-Dumpty effect, a phase dispersion arising from the uncertainty principle and related to the time-irreversibility in quantum theory. This effect resulted from the instability and imprecision of quantum operations \cite{Englert1988-ENGISC,Schwinger1988,Scully1989}.

Decoherence is one of the fundamental phenomena in physics. In a pioneering work, Zurek showed that decoherence explains the transition from quantum to classical phyics, wherein a system loses its quantum information through interactions with its environment \cite{Zurek1991}. 
Another type of the coherence loss know as dephasing where the randomization of the relative phases between quantum states occurs \cite{breuer2002}.
The density matrix for a two-level system is given by
$\rho_S (t)=\rho_{00}(t) \ket{0}\!\bra{0}{+}\rho_{01}(t)\ket{0}\!\bra{1}{+}\rho_{10}(t)\ket{1}\!\bra{0}{+}\rho_{11}(t)\ket{1}\!\bra{1}$ where $\ket{0}$ and  $\ket{1}$ represent the ground state and the excited state of the system, respectively. Through the interaction with its environment, the system can undergo decoherence via two primary mechanisms: amplitude damping and dephasing (or phase damping). In the amplitude damping process the energy is dissipated into the environment, leading to the decay of diagonal and off-diagonal elements as $\rho_{00}(t){=}1- \rho_{11}(t){=} 1-\rho_{11}(0) e^{-\gamma t}$, $\rho_{12}(t){=} \rho_{21}^{*}(t){=} \rho_{12}(0) e^{-\gamma t/2}$ where $\gamma$ is related to the relaxation time $T_1= 1/\gamma$. In contrast, dephasing primarily affects the quantum coherence without energy lost. Here, only the off-diagonal terms decay, evolving as $\rho_{12}(t){= }\rho_{21}^*(t){=} \rho_{12}(0) e^{-\gamma_\phi t}$, in which $\gamma_\phi$ is the dephasing rate responsible for the decay of $\rho_{12}(t)$
, related to dephasing time $T_\phi= 1/\gamma_\phi$ \cite{Manenti2023,Bertlmann2023,Yang_2016}
. 
The coherence decays exponentially with the transverse relaxation time $T_2= (\frac{1}{T_\phi}+\frac{1}{2T_1})^{-1}$.

In the quantum information processing (QIP) the essential requirements for implementing a practical quantum computer are  outlined by the DiVincenzo criteria \cite{DiVincenzo_2000}, and decoherence plays a significant role in several of these criteria.
In quantum dots (QDs) the dephasing caused by the lattice inertia leads to the strong fidelity limitations. In QDs, there are two primary types of coherence loss:  spin decoherence and charge dephasing that persists even at $T{=}0$ \cite{Khasanov2005,Jacak2008}.
In a SG interferometer, the  scattering of photons from a spin-carrying atom generates correlations between the system and its environment, which in turn lead to the decoherence of the atomic superposition state.
Furthermore, the relative acceleration between the interferomentric mass and the associated apparatus causes dephasing \cite{BERMAN1997}.

The emission of Bremsstrahlung through matter currents occurs when two spatially separated components of the state vector are brought together. 
The radiation field induces a loss of coherence, which can be effectively described by a gauge-invariant relativistic decoherence functional, representing a specific functional of the matter current densities. This decoherence functional is Lorentz covariant at finite temperatures and invariant at zero temperature, corresponding to the electromagnetic field vacuum \cite{breuer2001}.

Recent experimental efforts employing high-precision magnetic field gradients have successfully overcome the instability of quantum operations, allowing for the observation of spatial interference fringes with measurable phase stability \cite{Margalit_2021,Machluf2013}. In a full-loop SG interferometer, these magnetic gradients facilitate the entanglement of the spin and momentum. The process starts with a superposition state of spin-1/2 particles, typically created using a $\pi/2$ pulse. By applying a magnetic gradient pulse, the particles' momentum is split into two components, after which the two wavepackets are recombined in both position and momentum, achieving the desired entanglement.

Quantum field theory (QFT) techniques and the quantum Boltzmann equation (QBE) offer valuable insights into the mechanisms of decoherence and noise in open quantum systems. By investigating the interactions between the system and its environment, we can enhance our understanding of how the environment influences the system's evolution and the decay of quantum coherence. Such knowledge is essential for designing and controlling quantum systems, which plays a pivotal role in advanced quantum technologies, including quantum 
measurement and quantum sensing.
We have utilized the QBE technique to explore various phenomena in physics. For instance, we applied it to study the effects of cosmic microwave background radiation \cite{zarei2010,Bartolo:2019eac,Hoseinpour2020},
and calculate nuclear magnetic resonance (NMR) relaxation and decoherence times \cite{manshouri2021}. 
The QBE technique has demonstrated its versatility as a powerful tool for investigating a wide range of physical phenomena.
Additionally, we have put forward a novel experiment aimed at detecting axion-like particles (ALPs), as a spin-0 particle, absorbed by one of the atom interferometer's arms \cite{Hajebrahimi_2023}. In this proposed experiment, we use QBE formalism to describe the dephasing and phase shifts resulting from ALPs scattering off one "arm" of an atom interferometer.
Moreover, the absorption of a graviton, as a spin-2 particle, inducing decoherence in a spatial superposition of massive objects is considered in \cite{Sharifian_2024}. 
The decoherence effect results from the interaction of a system of particles with squeezed gravitational waves in an SG interferometer.

The decoherence caused by the Bremsstrahlung effect in a hypothetical interferometer device has previously been studied using the master equation with classical current of charged particles \cite{breuer2002}. However, to achieve more precise results, it is essential to consider all interactions of the matter field. Using QFT techniques in an open quantum system with the QBE approach offers significant advantages in evaluating all system interactions.
In this study, our objective is to apply QFT techniques using the QBE approach to study the interaction of fermions with photons, as a spin-1 particle, in the SG interferometer and compute the dephasing factor induced by the Bremsstrahlung effect.
To demonstrate the accuracy of our method, we will start by analyzing a particle with spin-1/2 in a superposition of spin-up and spin-down states. We will show that our results align with the decoherence functional of the classical Bremsstrahlung effect presented in Ref. \cite{breuer2001}. Additionally, by incorporating the spin term of the Hamiltonian, we will unveil the dependence of the results on velocity and time, which can increase the dephasing factor compared to the classical Bremsstrahlung, which lacks this significant dependence. 

In Section~\ref{sec:QBE}, we demonstrate that this method not only achieves the same result as presented in Subsection~\ref{sec:non-spin}, but it also reveals valuable insights into the dephasing factor caused by the Bremsstrahlung effect, as discussed in Subsection~\ref{sec:spin}. In Section~\ref{sec:dec} we connect our results to the measurable outcomes of atom interferometers. Finally, in Section~\ref{sec:con}, we present our conclusions based on the findings from the previous sections.


\section{Quantifying Decoherence Effects with QBE in Bremsstrahlung Emission Process}\label{sec:QBE}
Within the framework of open quantum systems, decoherence emerges from the Liouville-von Neumann equation describing a system interacting with its environment. The process is captured by the reduced density matrix $\rho_S$, where the decay of off-diagonal elements quantifies the loss of quantum coherence. 
Moreover, decoherence and dephasing can be characterized through quantum correlation functions $\braket{a_i(t) a^\dag_j(0)}$ ($a_i$ and $a^\dag_i$ are the annihilation and creation operators for state $i$).
In the case of QDs, the exciton single-particle correlation function governs the time evolution of the QD exiton state \cite{Jacak2008}. 
On the other hand, employing the influence phase functional in QED, a relativistic decoherence functional can be obtained, which provides a quantitative measure for the degree of decoherence of a quantum superposition.  In this method, dephasing factor is derived by integrating two-point correlation function $\braket{A_\mu(x)A_\nu(x')}$ along the closed path of the particle trajectory \cite{breuer2001}.

The QBE is a framework for studying the dynamics of quantum systems in contact with an environment. It allows us to calculate the time evolution of the density matrix of a quantum system coupled to an environment, taking into account the effects of decoherence and dissipation \cite{manshouri2021}. The QBE and the QFT techniques provide powerful tools for studying the effects of Bremsstrahlung radiation on quantum systems, and for understanding the underlying physical mechanisms that lead to decoherence.

In our two-level system with spin-1/2 particles, the off-diagonal elements of the density matrix $\rho^{(f)}_{ij}(t)$ provide information about the time evolution of the superposition state.  Here, we express the density matrix in terms of the number operators $\hat{\mathcal{D}}_{ij}(\mathbf{k})=b^\dag_i(\mathbf{k}) b_j(\mathbf{k})$ that describe the occupation of different quantum states with momentum $\mathbf{k}$. The expectation value of the number operator is given by
\begin{equation}
	\braket{\hat{\mathcal{D}}_{ij}(\mathbf{k})} = \text{Tr}(\hat{\rho}\hat{\mathcal{D}}_{ij}(\mathbf{k}))=(2\pi)^3 \delta^3(0) \rho^{(f)}_{ij}(\mathbf{k})~.
\end{equation}
The QBE can be expressed as a master equation derived under the assumption of Markovian dynamics, where the back-reaction of the environment on the system is negligible. In this regime, the system and environment remain uncorrelated throughout the time evolution. Physically, this corresponds to an environment so large that it cannot be influenced by the system, effectively acting as an unchanging, memoryless reservoir. In contrast, in non-Markovian dynamics, information can flow back from the environment to the system, giving rise to back-reaction effects \cite{zarei2021}.

Here, we focus on the Markovian case. In this regime, the time evolution of the density matrix elements is governed by a set of differential equations of the form \cite{Kosowsky:1994cy}
\begin{align}\label{boltzmann0}
	(2\pi)^3 \frac{k^0}{m_f} \delta^3(0)\frac{d}{dt}\mathcal{\rho}^{(f)}_{ij}(\mathbf{k},t)=i\braket{\l[H_{\textrm{int}}(t),\hat{\mathcal{D}}_{ij}(\mathbf{k})\r]}-\int_{0}^{t} dt' \braket{\l[H_{\textrm{int}}(t),\l[H_{\textrm{int}}^\dag(t'),\hat{\mathcal{D}}_{ij}(\mathbf{k})\r]\r]}~,
\end{align}
where $k^0$ is the energy of particles (specially fermions) with mass $m_f$, and
where $H_{\textrm{int}}$ is an effective interaction Hamiltonian that, in our case, is defined in terms of the first-order S-matrix and describes Bremsstrahlung process. The first term on the right-hand side of Eq. \eqref{boltzmann0} is referred to as the forward scattering term, while the second term is referred to as the damping or non-forward scattering term. In our calculation, the forward scattering term is expected to be zero. As a result, only the damping term will be taken into consideration. 
By including the collision term in the QBE to account for the interaction between the system and its environment, we can obtain a more realistic and accurate description of how decoherence occurs in the system.

Here, we are interested in the decoherence effect due to Bremsstrahlung radiation. 
The interaction Hamiltonian is given as the following form
\beq \label{Hint0}
H_{\textrm{int}}(t)= q_f\int d^3x\, \bar{\psi}(x)\gamma^{\mu}\psi(x)
A_{\mu}(x)~,
\eeq
where $\psi(x)$ is the fermion field, $A_{\mu}(x)$ is the electromagnetic field of photons, $q_f$ is the fermion's charge, and $\gamma^\mu$ is the Dirac $\gamma$-matrix. 
This interaction Hamiltonian describes the coupling between the fermion and the electromagnetic field, which can lead to the exchange of virtual photons and the emission and absorption of real photons. 

\begin{figure}
	\centering{\includegraphics[width=8cm]{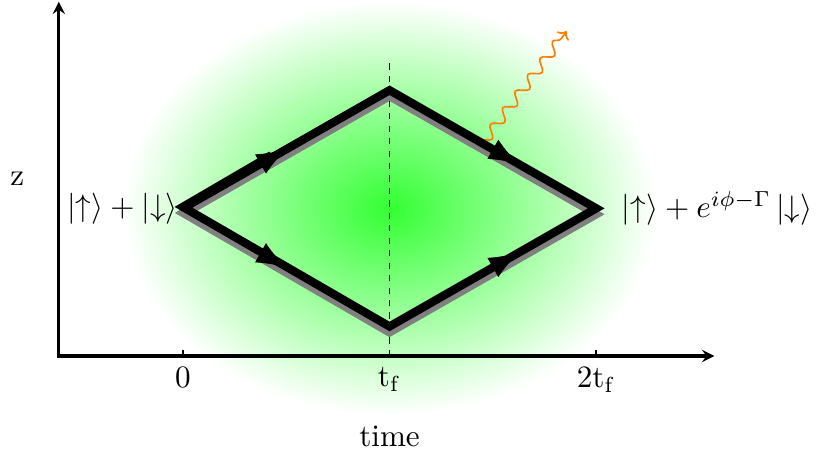}}
	\caption{The experimental setup involves a superposition of fermions interacting with Bremsstrahlung radiation. The thick black lines indicate the paths of the fermions. The initial state of the system is represented by $\ket{\uparrow}+\ket{\downarrow}$, while the final state is represented by $\ket{\uparrow}+e^{i\varphi-\Gamma t}\ket{\downarrow}$, where $\varphi$ is the induced phase and $\Gamma$ is the dephasing factor. The orange squiggly arrow indicates the emitted photon.}
	\label{fig:3}
\end{figure}

In the following, we will calculate the decoherence effect caused by both the spin and non-spin terms using Eq. \eqref{boltzmann0}. We  are interested in Bremsstrahlung emission process depicted in the Fig. \ref{fig:3}. 
The effective interaction Hamiltonian describing this process is given by
\beq \label{Hint1}
H_{\textrm{int}}(t)= q_f\int d^3x\, \bar{\psi}^-(x)\gamma^{\mu}\psi^+(x)A^-_{\mu}(x)~,
\eeq
where $\psi^+$ $(\bar{\psi}^-)$ and $A^+_\mu$ $(A^-_\mu)$ are linear in absorption (creation)
operators of fermions and photons respectively. Fourier transforms of the fields are defined using the following convention
\beq
\psi(x)=\psi^+(x)+\psi^-(x)~,
\eeq
where
\beq \label{fourierpsi}
\psi^+(x)=\int \frac{d^3p}{(2\pi)^3}\frac{m_f}{p^0}\sum_r\left[b_r(p)u_r(p)e^{-ip\cdot x}\right],
\eeq
and
\beq \label{fourierbarpsi}
\bar{\psi}^-(x)=\int \frac{d^3p}{(2\pi)^3}\frac{m_f}{p^0}\sum_r\left[b^{\dag}_r(p)\bar{u}_r(p)e^{ip\cdot x}\right]~,
\eeq
where $u_r$ is the free spinor solution of the Dirac equation with a spin index $r=1,2$. Here, $b_r(p)$ and $b_r^\dag(p)$ are the fermion annihilation and creation operators, respectively, which obey the canonical anti-commutation relation
\beq
\left\{b_r(p),b^\dag_{r'}(p')\right\}=(2\pi)^3\frac{p^0}{m_f}\delta^3(\p-\p')\delta_{rr'}.
\eeq
The expectation value of the creation and annihilation operators are given by
\bea
\braket{b^\dagger_{r'}(p')b_r(p)}=\frac{p^0}{m_f}(2\pi)^3 \delta^3(\mathbf{p}-\mathbf{p'}) \rho^{(f)}_{r' r}(\p)~,
\eea
and
\bea\label{expferm2}
\braket{b^\dagger_{r'_1}(p'_1)b_{r_1}(p_1)b^\dagger_{r'_2}(p'_2)b_{r_2}(p_2)}&=&\frac{p_1^0p_2^0}{m^2_f}(2\pi)^6 \delta^3(\mathbf{p_1}-\mathbf{p}'_1)\delta^3(\mathbf{p}_2-\mathbf{p}'_2) \rho^{(f)}_{r'_1 r_1} \rho^{(f)}_{r'_2 r_2}
\nonumber \\ && 
+\frac{p_1^0p_2^0}{m^2_f}(2\pi)^6 \delta^3(\mathbf{p}_1-\mathbf{p}'_2)\delta^3(\mathbf{p}_2-\mathbf{p}'_1) \rho^{(f)}_{r'_1 r_2}(\p_2) \left[\delta{r_1 r'_2}-\rho^{(f)}_{r'_2 r_1}(\p_1)\right]~.
\eea
The Fourier transform of the field of photons can be written as
\beq
A_\mu(x)=A^+_\mu(x)+A^-_\mu(x)~,
\eeq
where 
\begin{equation}
	A^+_\mu(x)= \int \frac{d^3q}{(2\pi)^3 ~ 2 q^0} \sum_s  a_s(q) \epsilon^s_{\mu}(q) e^{-iq x} ,
\end{equation}
and
\beq
A^-_\mu(x)= \int \frac{d^3q}{(2\pi)^3 ~ 2 q^0} \sum_s a^\dagger_s(q) \epsilon^{*s}_{\mu}(q) e^{iq x}~,
\eeq
that $\epsilon^s_{\mu}$ is the polarization vector of the photon and $a_s(q)$ and $a^\dag_s(q)$ are the annihilation and creation operators of the photon, respectively. These operators obey the canonical commutation relation
\begin{equation}
	\left[a_s(q),a^\dag_{s'}(q')\right]=(2\pi)^3 2 q^0 \delta^3(\mathbf{q}-\mathbf{q'})\delta_{ss'}~.
\end{equation}
The expectation value of the photon's creation and annihilation operators can be derived in a similar manner in the form of 
\begin{equation}
	\langle a^\dagger_{s'}(q')a_s(q)\rangle=2 q^0 (2\pi)^3 \delta^3(\mathbf{q}-\mathbf{q'}) \delta_{s' s} \frac{n^{(\gamma)}(\mathcal{\q})}{2}~,
\end{equation}
where $n^{(\gamma)}(\mathcal{\q})$ is the number density distribution of photons \cite{breuer2002}.
Now, plugging the Fourier transforms of the fields into Eq. \eqref{Hint1} we yield 
\beq \label{Hintfull}
H_{\textrm{int}}(t)= \int \frac{d^3p}{(2\pi)^3} \frac{p^0}{m_f}\frac{d^3p'}{(2\pi)^3}\frac{p'^0}{m_f}\frac{d^3q}{(2\pi)^3}\frac{1}{2q^0}\delta^3(\p+\q'-\q)\mathcal{M}\,e^{-i(p+q'-q)} b^\dag_{r'}(\q')b_r(\q)a(\p)
~,
\eeq
where
\beq \label{M0}
\mathcal{M}=\frac{q_f}{2 m_f} \bar{u}_{r'}(\mathbf{p'})\gamma^\mu u_r(\mathbf{p}) \epsilon^*_{s_\mu}(\mathbf{q})~,
\eeq
is the amplitude of the emission process.  
Using the Gordon Identity, the amplitude \eqref{M0} can be decomposed into two terms, the non-spin and spin terms
\beq
\mathcal{M}=\mathcal{M}_{\textrm{non-spin}}+\mathcal{M}_{\textrm{spin}}~,
\eeq
where the non-spin term is given by
\beq\label{non-spin-m}
\mathcal{M}_{\textrm{non-spin}}(pr,p'r',qs)=\frac{q_f}{2 m_f}(p'^{\mu}+p^{\mu}) \bar{u}_{r'}(\mathbf{p'}) u_r(\mathbf{p}) \epsilon^*_{s_\mu}(\mathbf{q})~,
\eeq
and the spin term is given by
\beq\label{spin-m}
\mathcal{M}_{\textrm{spin}}(pr,p'r',qs)=\frac{q_f}{2 m_f}i\sigma^{\mu\nu}(p'_{\mu}-p_{\mu}) \bar{u}_{r'}(\mathbf{p'}) u_r(\mathbf{p}) \epsilon^*_{s_\nu}(\mathbf{q})~,
\eeq
and 
the Dirac spinors is written in the following form
\bea \label{spinor}
u_r(\p) \simeq \left(\begin{array}{cc} \chi_r\\ \frac{\bm{\sigma}\cdot\p}{2m_f} \chi_r \end{array}\right)~,
\eea
in which $\chi_r$ is the two-component spinor. 

\subsection{Exploring the non-spin Term}\label{sec:non-spin}

We begin by focusing on the non-spin term, in which we substitute the non-spin part of the interaction Hamiltonian into the damping term given in Eq. \eqref{boltzmann0}.
The experimental arrangement illustrated in Fig. \ref{fig:3} entails interferometry employing spin-1/2 particles in the presence of a gradient magnetic field. Within this setup, particles with spin up and spin down experience distinct magnetic forces due to their different spin orientations. Consequently, the momentum in the z-direction, dependent on the parameter $r$, undergoes temporal variation and can be expressed as follows
\bea\label{momen}
\mathbf{k}(r) = \l(k_x , k_y , (-1)^r \l(-1+2 \theta(t-t_f)\r) k_z\r)~,
\eea
where $\theta(t)$ is the Heaviside function. The term $(-1)^r(-1+2\theta(t-t_f))$ accounts for the impact of spin on the momentum in the z-direction. The factor $(-1)^r$ signifies that the momentum is reversed for spin-up and spin-down particles, while $(-1+2\theta(t-t_f))$ switches the momentum's sign when time $t$ crosses the midpoint of the period, $t_f$.

In a closed loop interferometry device, a spin-1/2 particle is split into two spatially separated paths.
Using the momentum convention given in Eq. \eqref{momen}, the two paths of the interferometer can be described as follows:

Path 1: The spin-up particle travels in the positive z-direction in the first half-loop, with momentum $\mathbf{k}^{\uparrow,1} {=}(k_x, k_y, k_z)$, and then in the negative z-direction in the second half-loop, with momentum $\mathbf{k}^{\uparrow,2} = (k_x, k_y, -k_z)$.

Path 2: The spin-down particle travels in the negative z-direction in the first half-loop, with momentum $\mathbf{k}^{\downarrow,1}{ =} (k_x, k_y, -k_z)$, and then in the positive z-direction in the second half-loop, with momentum $\mathbf{k}^{\downarrow,2} = (k_x, k_y, k_z)$.

After taking the expectation value and integrating over the momenta, we arrive in the following expression for the damping term
\bea\label{coli}
\textrm{damping term}&=& ~ - \delta^3(0) \int_{0}^{t} dt'\int d^3 q \frac{n(\mathbf{q})}{ 4q^0}
\left [e^{-2iq^0t'}  \l(\delta_{r_1r_2} \delta_{r'_2i}\rho_{jr'_1}-\delta_{r'_2 i} \delta_{j r'_1}\rho_{r_1 r_2}\r) \right.  \nonumber  \\  
& & \left. \, +e^{2iq^0t'} \l(-\delta_{r_1 i} \delta_{j r_2}\rho_{r'_2 r'_1}+\delta_{j r_2} \delta_{r'_2 r'_1}\rho_{r_1 i}\r)\right ]\mathcal{M}(1)\mathcal{M}^\dag(2)
~,
\eea
where $\mathcal{M}(1)=\mathcal{M}(p_1r_1,p'_1r'_1,qs)$.

With the damping term known and using Eq. \eqref{non-spin-m}, one finds that the diagonal elements of the density matrix remain unchanged, $\rho_{ii}(t_f)= \rho_{ii}(0)$, demonstrates that Bremsstrahlung emission does not contribute to amplitude damping, in agreement with Ref. \cite{breuer2002}.
 The dephasing factor $\Gamma$ is determined by the off-diagonal elements of the density matrix at the final time $t_f$. This can be expressed using the following equation
\bea\label{eq:gamma}
\rho_{12}(t_f) &=& \rho_{12}(0) \,e^{-\Gamma(t_f)} ~,
\eea
where $\Gamma(t_f)$ is given by
\bea\label{gamma1}
\Gamma_{\textrm{non-spin}}(t_f) & =&
-\frac{\alpha}{\pi^2} \int^{\Omega_{\textrm{max}}}_0 d\omega \frac{n(\omega)}{\omega}  
\left[ \l(1-\cos\l(\frac{\omega t_f}{2}\r)\r)-\frac{1}{4}\l(1-\cos\l(\omega t_f\r)\r)\right] \nonumber \\ 
&& \:\:\: \:\:\:  \:\:\:  \:\:\:  \:\:\:  \:\:\:  \:\:\:  \:\:\:   \times  \int d\Omega(\hat{q})\left[\frac{1-v^2}{(1+\hat{q}.\mathbf{v})^2}+\frac{1-v^2}{(1-\hat{q}.\mathbf{v})^2}-2 \frac{1+v^2}{1-\l(\hat{q}.\mathbf{v}\r)^2}\right]~,
\eea
where $\mathbf{v}=(\mathbf{k}/m)$ is the velocity vector of particles.
If we express the photon density distribution as $n(\mathbf{q})=(n(\mathbf{q})-1)+1$, we can decompose $\Gamma$ into two factors: $\Gamma_{\textrm{vac}}$, which is proportional to a factor of 1 and corresponds to vacuum {dephasing}, and $\Gamma_{th}$, which is proportional to a factor of $n(\mathbf{q})-1$ and corresponds to thermal {dephasing}. By using the photon density distribution, given by
\beq
n(\omega)=\coth\left(\frac{\beta \omega}{2}\right),
\eeq
we can obtain the same result as Eqs. \eqref{clsvac} and \eqref{clsth}, which was derived in Appendix \ref{Decoherence} using the influence phase functional method, in the limit where $|\mathbf{q}| \ll |\mathbf{k}|$ \cite{breuer2001}.

If we use a fixed maximal spatial distance $\xi=vt_f$, the dephasing factor in Eqs. \eqref{clsvacm} and \eqref{clsthm} in the nonrelativistic condition can be written as
\begin{equation}\label{non-spin-vac}
	\Gamma_{\textrm{non-spin-vac}}(t_f)= -\frac{2 \alpha}{\pi} \ln \l(g\, \Omega_{\mathrm{max}} t_f \r) \frac{\xi^2}{t_f^2}~,
\end{equation}
and
\begin{equation}\label{non-spin-ter}
	\Gamma_{\textrm{non-spin-th}}(t_f)= -\frac{8 \alpha}{3\pi}\l[ \ln\l(\frac{\sinh(t_f \pi/ \beta)}{(t_f \pi / \beta)}\r)-\frac{1}{4}\ln\l(\frac{\sinh(2 t_f \pi/ \beta)}{( 2 t_f \pi/ \beta)}\r)\r] \frac{\xi^2}{t_f^2}~.
\end{equation}
In the interaction Hamiltonian given in Eq. (\ref{Hintfull}), the emission or absorption of a real photon corresponds to a measurement of the photon's state, which collapses the state of the fermion field into a particular eigenstate. This can be understood by considering the emission of a real photon, for example. In this case, the initial state of the system is a coherent superposition of different eigenstates of the fermion's Hamiltonian, which includes the possibility of the fermion having absorbed a virtual photon from the external electromagnetic field. However, when a real photon is emitted, the system's state is projected onto a particular eigenstate that corresponds to the final state of the fermion and the emitted photon. This projection collapses the coherence between different eigenstates, as the probability amplitudes corresponding to different eigenstates are no longer in a coherent superposition.

\subsection{Exploring the spin term}\label{sec:spin}

To this end, we consider the spin part of the emission amplitude Eq. \eqref{Hint1}
and we will show it can be noticeable in the current SG interferometers, although this term is claimed to be negligible in Ref.\cite{breuer2001} due to large length scales.
In contrast to the non-spin term, which yields purely real values of the exponential factor in Eq.~\eqref{eq:gamma}, the spin-dependent term contributes both real and imaginary parts
	\begin{equation}
		\rho_{12}(t_f) = \rho_{12}(0)\,\exp\big(-\Gamma - i\varphi\big).
	\end{equation}
	The real component $\Gamma$ corresponds to the dephasing factor, leading to suppression of the interference terms. In contrast, the imaginary contribution $i\varphi$ represents a phase shift, which distorts the interference pattern without reducing its overall visibility \cite{breuer2002,breuer2001}.
We use Eq. \eqref{coli} and Eq. \eqref{spin-m} to calculate off-diagonal elements of the density matrix, we can express $\Gamma$ as follows:
\bea\label{gamma21}
\Gamma_{\textrm{spin-vac}}(t_f) &=& -\frac{\alpha}{2\pi^2}\frac{1}{m_f} \int_{0}^{\Omega_{\max}} d\omega \int d\Omega(\hat{q}) v \cos(\theta) \l(\frac{1}{1-(v \cos\theta)^2}\r)^2 \nonumber \\ 
&& \times \l[8\sin(\omega t_f)\sin^3(v t_f \omega \cos\theta)-2\cos(2 t_f\omega)\sin(2 v t_f \omega \cos\theta)+\sin(4 t_f v \omega \cos\theta)\r]~,
\eea
\bea\label{gamma22}
\Gamma_{\textrm{spin-th}}(t_f) &=& -\frac{\alpha}{2\pi^2}\frac{1}{m_f} \int_{0}^{\Omega_{\max}} d\omega \int d\Omega(\hat{q}) v \cos(\theta) \l(n(\omega)-1\r) \l(\frac{1}{1-(v \cos\theta)^2}\r)^2 \nonumber \\ 
&& \times \l[8\sin(\omega t_f)\sin^3(v t_f \omega \cos\theta)-2\cos(2 t_f\omega)\sin(2 v t_f \omega \cos\theta)+\sin(4 t_f v \omega \cos\theta)\r]~,
\eea

where we can decompose $\Gamma_{\textrm{spin}}$ to $\Gamma_{\textrm{spin-vac}}$ and $\Gamma_{\textrm{spin-th}}$ by writhing the photon density distribution in the form of $n(\mathbf{q})=(n(\mathbf{q})-1)+1$. Also we drive the imaginary term as
\bea\label{phase}
i\varphi_{\textrm{spin-vac}}(t_f) &=& -i\frac{\alpha}{2\pi^2}\frac{1}{m_f} \int_{0}^{\Omega_{\max}} d\omega \int d\Omega(\hat{q})~ v^2 \cos(\theta) n(\omega) \frac{1}{1-(v \cos\theta)^2}\nonumber \\ 
& &\times \l[\sin(\omega t_f)\sin( v t_f \omega \cos\theta)-\sin(2 \omega t_f)\sin(2 v t_f \omega \cos\theta)+\sin(3 v t_f \omega \cos\theta)\r]~.
\eea
This term arises from the combined contributions of the non-spin and spin parts in Eqs.~\eqref{non-spin-m} and \eqref{spin-m}. Its amplitude carries an additional factor of $v/c$ relative to the dephasing factor $\Gamma_{\textrm{spin}}$~. 

\begin{figure}	
	\centerline{\includegraphics[width=10cm]{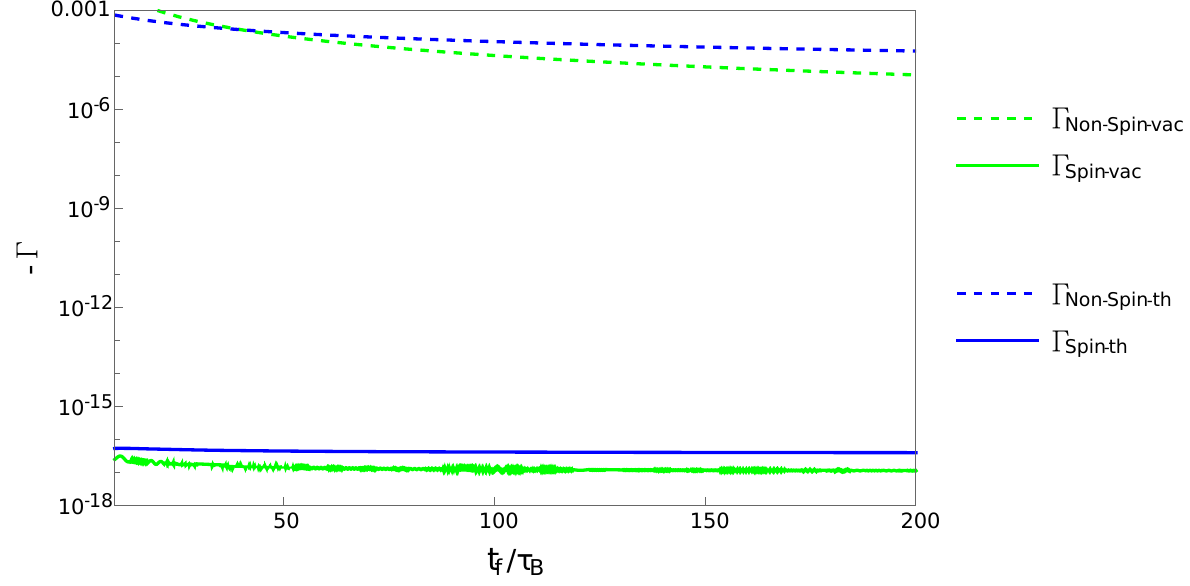}}
	\caption{The vacuum and thermal (at the temperature $1^{\circ}K$) dephasing factors of non-spin term (dashed lines) and spin term (solid lines) for the fixed maximum spacial distance $\zeta=O(10^{-3}m)$ of electrons is plotted. The dephasing factor of the spin term $\Gamma_{\textrm{spin-vac}}$ and $\Gamma_{\textrm{spin-th}}$ is calculated numerically.}
	\label{fig:4}
\end{figure}

In Fig. \ref{fig:4} the vacuum and thermal decoherence of non-spin and spin terms of electrons are shown where the spin term is calculated numerically. It is noticeable that for the fixed maximum spacial distance $\zeta \approx O(10^{-3}m) $ the decoherence induced by the spin term (Eqs. \ref{gamma21} and \ref{gamma22}) is negligible relative to the non-spin term (Eqs. \ref{non-spin-vac} and \ref{non-spin-ter}) in consistence with the results and prediction about these terms in Ref. \cite{breuer2001}. 
Notably, analogous behavior has been observed in QD systems, in which the magnon-assigned decoherence at  $T=0$ disappears in compare with the phonon-induced QD charge dephasing that exists at this condition \cite{Jacak2008}.

\section{Dephasing factor for Atom Interferometry}\label{sec:dec}

By employing the QBE and QFT techniques to derive the dephasing factor arising from Bremsstrahlung emission, we uncover insights that hold significant relevance for various applications in atom interferometry. These techniques offer a powerful means to explore and recover microscopic details of atom interferometry's interaction with weak external fields, opening up avenues for investigating exotic particles like axions and probing the intricate nature of gravitational waves. 

Atom interferometers stand out as exceptionally promising devices for scientific exploration due to their essential properties. One of their distinctive features is the absence of a minimum energy threshold, making them ideal for ultra-sensitive applications \cite{rudolph2020}. Additionally, these interferometers display remarkable adaptability across a diverse spectrum of momentum and energy values, allowing them to accommodate various types of test particles and enabling versatile experimental setups.

In atom interferometers, decoherence plays an important role as a crucial quantum observable, influencing the amplitude of the interference fringe \cite{Riedel_2013}. 
Furthermore, by measuring the relative phase between two distinct measurements, atom interferometers enable the determination of quantum observables, such as the phase shift resulting from the path differences between the two arms. 

In the context of atom interferometry, the visibility ($\mathcal{V}$) is a crucial parameter used to quantify the degree of decoherence in the system. As the system undergoes decoherence, due to various processes, such as the emission of Bremsstrahlung radiation, the interference fringes become less distinct, leading to a reduction in visibility.
The visibility is related to the dephasing factor ($\Gamma$) through the expression of \cite{margalit2018}, 
\bea
\mathcal{V} = e^{-\Gamma}. 
\eea
Measuring the visibility is crucial for optimizing the performance of atom interferometers and improving their sensitivity for various applications, such as detecting dark matter, gravitational waves, and other fundamental physical phenomena.

Neutral two-level atoms, such as ${}^{87}\mathrm{Rb}$ \cite{margalit2018,Du_2022,Aveline2020}, $\mathrm{Nb}$ \cite{Pino_2018}, and Carbon \cite{Margalit_2021}, are the prevalent choice in the majority of atom interferometers. The creation of this two level system experimentally  be achieved via the non-linear Zeeman effect to create two energy levels $\ket{2,2}$ and $\ket{2,1}$ \cite{Elliott_2018}. Practically, by applying a radio-frequency $\pi/2$ pulse we create a superposition of Dirac spinors Eq. \eqref{spinor}, and by a magnetic gradient pulse we impose a different magnetic potential to create two atomic spatial states with different momentum. 
These atoms are highly practical and widely used in experimental setups due to their ease of manipulation, long coherence times, and well-understood atomic properties. 

Investigating inhomogeneity in the atomic interferometer is intriguing from two perspectives. Firstly, it allows us to explore the impact of decoherence radiation on the visibility function and identify crucial parameters influencing the system. Secondly, this formalism presents an opportunity to study the decoherence effects induced by other weak fields, such as axions or gravitational waves. The versatile nature of this approach opens up new possibilities for understanding and controlling decoherence in a variety of quantum systems, offering potential applications in the realm of precision measurements and quantum sensing.

\begin{figure}
	\includegraphics[width=.45\linewidth]{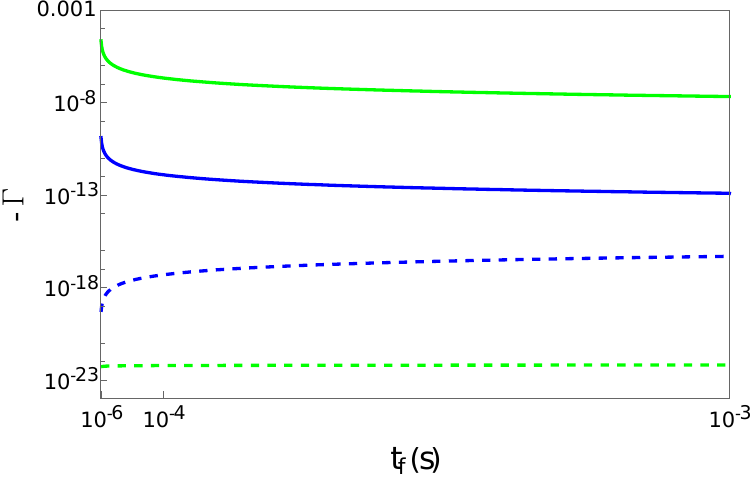}	
	\put(-200,150){(a)}	
	\includegraphics[width=.60\linewidth]{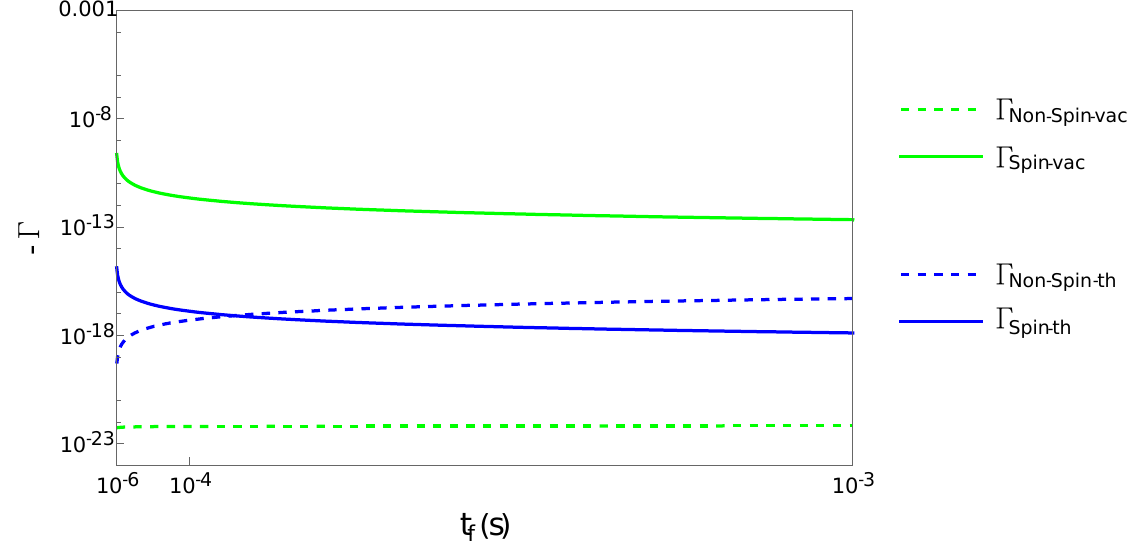}
	\put(-275,150){(b)} 
	\caption{The plot illustrates the vacuum and thermal dephasing factor (at the temperature $1~\mathrm{K}$) of the non-spin term (dashed lines) and spin term (solid lines) for a fixed velocity $v/c=10^{-11}$ for electrons in (a) and Ag atoms in (b).
	} 
	\label{fig:main}
\end{figure}

In Fig. \ref{fig:main}(a), we present the results obtained from the SG interferometer experiment \cite{margalit2018,Margalit_2021}, where we consider a flight time from splitting to recombination ranging from $10^{-6} s$ to $10^{-3} s$ with a fixed splitting velocity of approximately $v/c \approx 10^{-11}$. Under these conditions, the decoherence induced by the vacuum spin term $\Gamma_{\textrm{spin-vac}}$ exceeds the decoherence induced by the vacuum non-spin term $\Gamma_{\textrm{non-spin-vac}}$ by more than 15 orders of magnitude. This significant difference highlights the potential impact of vacuum spin decoherence on the accuracy of recent experiments. It is crucial to note that the vacuum spin dephasing factor is inversely related to the test particles' velocity, leading to its pronounced effect in the system.


As we can see, while the decoherence caused by the non-spin term (Eqs. \ref{non-spin-vac} and \ref{non-spin-ter}) is negligible in atom interferometers, the spin term (Eqs. \ref{gamma21} and \ref{gamma22}) becomes prominent due to the low atomic velocity, leading to discernible decoherence effects. It is crucial to account for these factors in experimental setups to ensure the reliability and precision of atom interferometry measurements.

On the other hand, to have an efficient comparison, in Fig. \ref{fig:main}(a) the electron mass is considered as the test particle. However, the dephasing factor of spin term depends on the inverse of the particle's mass. In interferometer devices working with neutral particles , the result that has been shown in Fig. \ref{fig:main}(b) for spin term decreases up to 5 orders, while the non-spin term dephasing factor does not have any dependency on the test mass (Eqs. \ref{clsvac} and \ref{clsth} ).
It is worth mentioning, the vacuum phase factor $\Phi_{\textrm{spin-vac}}$  calculated by Eq.\eqref{phase} in these conditions are negligible in comparison with the maximum sensitivity of recent sensors \cite{Pino_2018}.

\section{ Conclusion}\label{sec:con}

In this study, we have investigated bremsstrahlung emission from a microscopic perspective. This effect is a potential noise source that can contribute to the degradation of SG atom interferometry. 
We used QBE to investigate vacuum- and thermally-induced decoherence caused by the emission of Bremsstrahlung. By considering the spatial superposition of two distinct spin states in an interferometer device, we explored the loss of coherence by deriving the time evolution of the density matrix components. Our findings are consistent with the results obtained using the influence phase functional method described in Refs. \cite{breuer2001,breuer2002}.
We also took into account the spin term that appears in the interaction Hamiltonian to calculate the induced decoherence. In this case, as well as the dephasing factor, a phase shift is also induced, although it is very small for electrons and atoms.

Our findings show a noticeable difference between the dephasing factor induced by Bremsstrahlung effect in the non-spin \eqref{gamma1} and the spin terms \eqref{gamma21} and \eqref{gamma22} at low velocities ($v/c \approx 10^{-11}$), indicating that the spin term decoherence in the vacuum condition $\Gamma_{\textrm{spin-vac}}$ exceeds the non-spin decoherence in this condition by more than 15 orders of magnitude. The observed behavior is attributed to the inverse dependency of the dephasing factor on the relative velocity of the test particles along the respective paths. Also, this loss of coherence induced by the spin term depends on the mass of falling particles inversely. Consequently the use of massive particles in atom interferometers results in a diminished impact of Bremsstrahlung effect on the visibility of the fringes in comparison with light particles as considered in Ref. \cite{breuer2001}.

This result highlights an intrinsic source of noise in atom interferometers. Similar degradation may also arise from other factors, including environmental noise, imperfect experimental control, and interactions with surrounding particles. As the sensitivity and accuracy of atom interferometers improve in the search for new physics, they will inevitably face coherence loss induced by intrinsic decoherence effects, such as Bremsstrahlung emission—an unavoidable noise in the experimental setup.
This source of decoherence adds to other well-known noise-inducing phenomena, including non-inertial jitter and gravity gradient noise \cite{Toro2021,Marshman_2020}; quantum projection noise from measurements of uncorrelated two-level systems \cite{Doring2010}; and Dicke noise in optical clocks \cite{Westergaard_2010}.
	
	Moreover, our calculation of Bremsstrahlung emission via spin-1 particles advances the understanding of coherence loss in SG interferometers. This work complements earlier studies of decoherence induced by spin-0 and spin-2 particles interacting with fermions in SG interferometers \cite{Hajebrahimi_2023,Sharifian_2024}. To complete the picture, it would be valuable to investigate decoherence arising from Bremsstrahlung emission involving spin-1/2 particles, such as neutrinos, which we intend to explore in future work.

\begin{acknowledgments}
	
	M. Z. would like to thank INFN and department of Physics and Astronomy “G. Galilei” at University of Padova and also the CERN Theory Division for warm hospitality while this work was done.

\end{acknowledgments}

\appendix
\section{Decoherence function from Bremsstrahlung Emission using Influence Functional Method}\label{Decoherence}

In this section, we provide an overview of the relativistic formulation for the loss of coherence resulting from Bremsstrahlung emission, as developed in \cite{breuer2001}. We begin by introducing the influence phase functional, which characterizes the phase acquired by a quantum system through its interaction with the surrounding environment. In the context of Bremsstrahlung emission, the environment corresponds to the electromagnetic field generated by charged particles.

The interaction Hamiltonian density between the matter current density $j^\mu(x)$ and the transverse radiation field $A^\mu(x)$ is expressed as follows
\beq \label{H}
\mathcal{H}_I(x)=j^\mu(x)A_\mu(x)~,
\eeq
The condition $\partial_\mu j^\mu =0$ is satisfied by $j^\mu(x)$. The objective is to derive an exact representation of the reduced density matrix $\rho_m$ pertaining to the matter degrees of freedom by eliminating the variables associated with the electromagnetic radiation field. This is accomplished through the following formal equation, which establishes a connection between the density matrix $\rho_m(t_f)$ of the matter at a final time $t_f$ and the density matrix $\rho(t_i)$ of the combined matter-field system at an initial time $t_i$
\beq
\rho_m(t_f)=\mathrm{Tr}_f \left\{ T_\leftarrow \exp \left[\int_{t_i}^{t_f} d^4 x \mathcal{L}(x) \right ]\rho(t_i) \right\}~,
\eeq
where $\rho(t_i)$ is the initial state of the combined matter-field system. One can consider an initial state of low entropy which is given by a product state of the form
\beq
\rho(t_i)=\rho_m \otimes \rho_f~, 
\eeq
where $\rho_m(t_t)$ is the density matrix of the matter at the initial time and the density matrix $\rho_f$
of the radiation field describes an equilibrium state at temperature $T_{\textrm{em}}$ that can be write as
\beq
\rho_f=  \frac{\exp \l (-\beta H_{f} \r)}{\textrm{tr}_f [\exp \l (-\beta H_{f} \r)]}~,
\eeq
where $H_f$ is the free Hamiltonian for the radian field and $\beta=1/(k_\textrm{B}T_{\textrm{em}})$ is the Boltzmann factor. 
Here, the time-ordered exponential of the Liouville superoperator $\mathcal{L}(x)$ acts on density operators $\rho$ of the combined matter-field system and is taken from the initial time $t_i$ to the final time $t_f$, and the trace is taken over the variables of the electromagnetic radiation field.
The Liouville superoperator $\mathcal{L}(x)$ is defined in terms of the Hamiltonian density $\mathcal{H}$ of the combined matter-field system as follows
\beq
\mathcal{L}(x)\rho=-i[\mathcal{H},\rho]~.
\eeq
A charged particle, such as an electron, is emitted from a source and has two distinct paths, denoted as $y_1$ and $y_2$, leading to an observation of an interference pattern on a screen. To quantify the degree of decoherence, which arises when the coherence between the two probability amplitudes is lost due to interactions with the environment, their proposal suggests utilizing this specific physical setup as a prototypical interference device. The two wave packets associated with the two paths are represented as $\ket{\Psi_1(t_i)}$ and $\ket{\Psi_2(t_i)}$, describing the respective probability amplitudes, and their coherent superposition can be denoted by the wave function
\beq \label{Psiti}
\ket{\Psi(t_i)}=\ket{\Psi_1(t_i)}+ \ket{\Psi_2(t_i)}~.
\eeq
The representation of the electron state using the density matrix $\rho_m(t_i)$ offers a valuable approach to study quantum systems. The density matrix consists of four elements: $\rho_m(t_i) = \rho_{11}(t_i) + \rho_{12}(t_i) + \rho_{21}(t_i) + \rho_{22}(t_i)$, where $\rho_{ij}(t_i) = \ket{\Psi_i(t_i)}\bra{\Psi_j(t_i)}$.
The interference term $\rho_{12}(t_i) + \rho_{21}(t_i)$ accounts for the probability of the electron being in a superposition of two states. Our objective is to investigate the structure of the electron density matrix $\rho_m$ in the presence of an electromagnetic radiation field using an influence superoperator. This superoperator allows us to consider the effects of the environment on the system.


The characteristics of a low-energy regime in which the frequency of the emitted radiation is much smaller than the rest mass of the electron (i.e., $\hbar \omega \ll mc^2$). In this regime, the pair creation and annihilation amplitudes can be neglected, and the matter current density can be treated as a given classical field. 
When the matter's current density can be treated as a classical current, it means that the radiation emitted by the currents is much larger in wavelength compared to the Compton wavelength of the electron. 
The approximation of classical currents is justified when the wavelength of the radiation emitted by the currents is much larger than the Compton wavelength of the electron. This condition is expressed as $\bar{\lambda}\gg \bar{\lambda}_C$, where $\bar{\lambda}$ is the wavelength of the radiation and $\bar{\lambda}_C$ is the Compton wavelength of the electron. 

The second condition is that the motion of the electric current can be reasonably described within a semiclassical approximation. This condition imposes a requirement that the velocity uncertainty $\Delta v$ should be significantly smaller than the typical velocity $v$, i.e., $\Delta v/v\ll 1$. Here, $v$ represents the average velocity of the electrons in the current, while $\Delta v$ refers to the uncertainty in their velocities, which arises from the quantum mechanical uncertainty principle. 
This condition is necessary for the semiclassical approximation to be valid because it allows us to treat the electrons as particles with well-defined trajectories, while still taking into account their wave-like properties.

\begin{figure}[t]
	\centerline{\includegraphics[width=7cm]{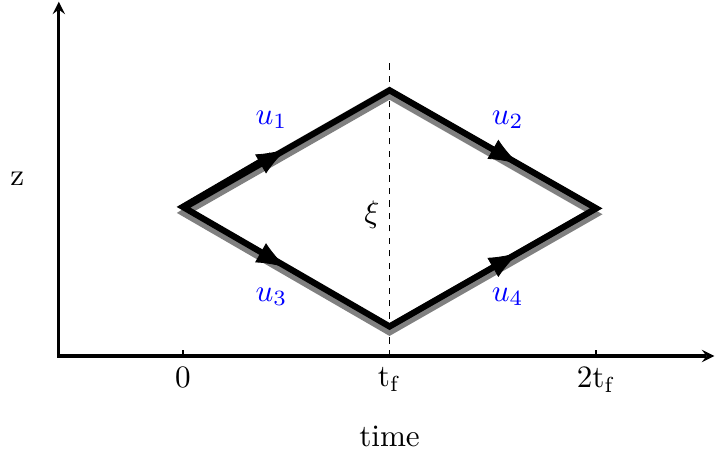}}
	\caption{The loop consists of two paths, each comprising four straight world-line segments with 4-velocities of $u_1$, $u_2$, $u_3$, and $u_4$. The loop spans a time period of $2t_f$, and the maximum spatial separation between the two paths, denoted as $\xi$, occurs at time $t_f$.}
	\label{fig:2}
\end{figure}

Based on these semiclassical conditions, one can now assume that $\rho_m(t_i)$ represents an approximate eigenstate of the current density. It can be supposed that current density operator $j^\mu(x)$ is approximately equal to a classical current density $s^\mu(x)$ acting on the same state $\ket{\Psi(t_i)}$
\beq
j^\mu(x)\ket{\Psi(t_i)} \approx s^\mu(x)\ket{\Psi(t_i)}~.
\eeq
In the other word, the expectation value of the current density is given by
\bea
\braket{j^\mu(x)} = Tr_m\left\{j^\mu(x)\hat{\rho}_m(t_i)\right\} = s^\mu (x)~.
\eea
In the case of the interference device described above, the superposition \eqref{Psiti} consists of two current eigenstates
\bea\label{current} 
j^\mu(x)\ket{\Psi_1(t_i)} &\approx& s^\mu_1 (x) \ket{\Psi_1(t_i)}~, \nonumber\\
j^\mu(x)\ket{\Psi_2(t_i)} &\approx& s^\mu_2 (x) \ket{\Psi_2(t_i)}~.
\eea
Furthermore, these currents are assumed to be concentrated within two world tubes around the paths $y_1$ and $y_2$ of the interference device, respectively as they are depicted in Fig. \ref{fig:2}.

It is worth noting that the approximation sign in Equations \eqref{current} indicates that they are valid only in the limit where the quantum state is localized within the world tubes around the classical paths.

The influence phase functional is a mathematical expression that characterizes the effect of an environment on a quantum system. The influence phase functional captures the loss of coherence induced by the environment, which can lead to a decay of quantum superpositions and the emergence of classical behavior. It is derived from the reduced density matrix, which describes the state of the system after tracing out the degrees of freedom of the environment. It is given by $\rho_m(t_f)$ the reduced density matrix of the system at the final time $t_f$ as the following form
\bea
\rho_m(t_f) \approx \rho_{11}(t_i) + \rho_{22}(t_i) + \exp(i\Phi)\rho_{12}(t_i) + \exp(-i\Phi^*)\rho_{21}(t_i)~,
\eea
where the exponential factors $\exp(i\Phi)$ and $\exp(-i\Phi^*)$ represent the dephasing factor induced by the emission of Bremsstrahlung on the off-diagonal elements of the reduced density matrix. 
The dephasing factor $\exp(i\Phi)$ is given as the following form
\bea\label{Phi0}
\exp(i\Phi)&=& A[s_1] A[s_2]^*\exp \left[ \frac{1}{2} \int d^4 x \int d^4 x' D_-(x-x')_{\mu \nu} s^\mu_1 (x) s^\nu_2 (x') \right. \nonumber\\
&& \left.  ~~~~~~~~~~~~~~~~~~~~~~~+\frac{1}{2} \int d^4 x \int d^4 x' D_+(x-x')_{\mu \nu} s^\mu_2 (x) s^\nu_1 (x')\right]~, 
\eea 
where $D_+(x-x'){\mu \nu}$ and $D_-(x-x')_{\mu \nu}$ are two-point correlation functions defined as
\bea\label{Dminusplus}
D_+(x-x')_{\mu \nu} = \braket{A_\mu(x)A_\nu(x')} , \nonumber\\
D_-(x-x')_{\mu \nu} = \braket{A_\nu(x') A_\mu(x)}~,
\eea
and  
\bea\label{}
A[j] = \exp \l[
- \frac{i}{2} \int d^4 x \int d^4 x' D_\textrm{F}(x-x')_{\mu \nu} j^\mu (x) j^\nu (x')\r]~,
\eea 
is a functional integral over all possible configurations of the electromagnetic
field $A_\mu(x)$ that gives the probability amplitude for the vacuum state to transition to itself in the presence of the classical current density $j^\mu(x)$~, and $D_\textrm{F}(x-x')$ is the Feynman propagator
\begin{equation}
	iD_\textrm{F}(x-x')_{\mu \nu} = \braket{T_\leftarrow\left[A_\mu(x)A_\nu(x')\right]}~.
\end{equation}

At zero temperature $A[j]$ is the vacuum-to-vacuum amplitude in the presence of a classical current density $j^\mu(x)$.
The first term on the right-hand side of Eq.\eqref{Phi0} corresponds to the product of the vacuum-to-vacuum amplitudes in the presence of $s_1$ and $s_2$ and represents virtual processes involving the emission and reabsorption of photons by either $s_1$ or $s_2$. 
Correspondingly, the exponential on the right-hand side of the above equation is the contribution of the emission of at least one photon.
These virtual processes also contribute to the decoherence functional. In this case, the environment is the radiation field that is emitted and absorbed by the currents $s_1$ and $s_2$. The emitted photons carry away information about the path taken by the electron and contribute to the loss of coherence.
Furthermore, at finite temperatures, thermally induced emission and absorption processes occur,
in addition to the virtual processes described above. The radiation field produced by these processes has the typical spectrum of Bremsstrahlung.

The exponential part of Eq. \eqref{Phi0} which contributes to the dephasing factor can be written as
\beq
\Gamma[c]=-\frac{i}{2}\int d^4x \int d^4 x' D_\textrm{F}(x-x')_{\mu\nu}c^{\mu}(x)c^{\nu}(x') + \textrm{c.c.}~,
\eeq
where $c^\mu(x)=[s_1^\mu(x)-s_2^\mu(x)]/\sqrt{2}$ is the current difference.

In the interferometer device, the electron beam split and recombined to produce an interference pattern. The device consists of a closed loop made up of two paths, each consisting of four straight world-line segments with 4-velocities of $u_1$, $u_2$, $u_3$, and $u_4$, occurring over a period of time $2 t_f$.
Assuming symmetry, the loop consists of four straight world-line segments, with vertices denoted by $a_0$, $a_1$, $a_2$, $a_3$, and corresponding four velocities of $u_1$, $u_2$, $u_3$, $u_4$ (Fig. \ref{fig:2}). We further assume that the arrangement is symmetric, that is, $u_1=u_4$ and $u_2=u_3$, also $a_1-a_0=a_2-a_3$ and $a_2-a_1=a_3-a_0$.
With the help of this arrangement and some calculations, one can find the decoherence functional
\bea\label{Gamma0}
\Gamma(t_f)= -\frac{\alpha}{(\pi)^2} \int_{0}^{\Omega_{\max}} \frac{d\omega}{\omega} \coth\left(\frac{ \beta \omega}{2}\right)\l[\left(1-\cos\left(\omega t_f\right)\right)-\frac{1}{4}(1-\cos(2\omega t_f))\r]\int d\Omega(\hat{q})\,\omega^2\l[\frac{u_2}{q\cdot u_2}-\frac{u_4}{q\cdot u_4}\r]^2~,
\eea
where the photon's four-momentum is $q = (\omega,q\hat{q})$.
In order to calculate the decoherence functional one can decompose Eq.\eqref{Gamma0}
into a vacuum contribution and a thermal contribution which vanishes for $T_{\textrm{em}}=0$.  
After taking the frequency and $d\Omega(\hat{q})$ integrals one finds the vacuum
and the thermal contributions to the decoherence functional as the following form respectively 
\begin{equation}\label{clsvac}
	\Gamma_{\textrm{vac}}(t_f)= -\frac{6 \alpha}{\pi} \ln \l(g\, \Omega_{\mathrm{max}} t_f \r) \l(\frac{1}{2 v}\tanh^{-1}(2 v)-1\r)~,
\end{equation}
and
\begin{equation}\label{clsth}
	\Gamma_{\textrm{th}}(t_f)= -\frac{8 \alpha}{\pi}\l[ \ln\l(\frac{\sinh(t_f  / \tau_B)}{t_f /\tau_B}\r)-\frac{1}{4}\ln\l(\frac{\sinh(2 t_f/ \tau_B)}{2 t_f / \tau_B}\r)\r] \l(\frac{1}{2 v}\tanh^{-1}(2 v)-1 \r)~,
\end{equation}
where $\ln(g) \approx 0.577$ is Euler's constant, $\alpha$ is the fine-structure constant,  $\tau_B=\pi/\beta$ and $v$ is the velocity of particles in the z-direction. Actually, we use the fact that Eq. \eqref{Gamma0} is Lorentz invariance, so we can calculate in the frame where $\vec{v}=(0,0,v)$.  
One can also expand the $\tanh^{-1}$ function and write expressions for $\Gamma_{\textrm{vac}}$ and $\Gamma_{\textrm{th}}$
in the non-relativistic limit as the following form
\begin{equation}\label{clsvacm}
	\Gamma_{\textrm{vac}}(t_f)= -\frac{2 \alpha}{\pi} \ln \l(g \,\Omega_{\mathrm{max}} t_f \r) \frac{\xi^2}{t_f^2}~,
\end{equation}
and
\begin{equation}\label{clsthm}
	\Gamma_{\textrm{th}}(t_f)= -\frac{8 \alpha}{3\pi}\l[ \ln\l(\frac{\sinh(t_f / \tau_B)}{t_f / \tau_B}\r)-\frac{1}{4}\ln\l(\frac{\sinh(2 t_f / \tau_B)}{(2 t_f  / \tau_B}\r)\r]  \frac{\xi^2}{t_f^2}~,
\end{equation}
where $\xi$ is a fixed maximal spatial distance between the paths.

\bibliography{reference-decoherence}
\end{document}